\documentclass[a4paper,fleqn,usenatbib]{mnras}

\usepackage[T1]{fontenc}
\usepackage{ae,aecompl}

\usepackage{graphicx}	
\usepackage{amsmath}	
\usepackage{amssymb}	%

\newcommand{\st}{^{\circ}}

\title[SAGE non-convex model of 809 Lundia]{A new non-convex model of the binary asteroid (809) Lundia obtained with the SAGE modelling technique}

\author[P. Bartczak et al.]{
P. Bartczak$^{1}$\thanks{E-mail:przebar@amu.edu.pl},
A. Kryszczy{\'n}ska $^{1}$,
G. Dudzi{\'n}ski$^{1}$,
M. Poli{\'n}ska$^{1}$,
F. Colas$^{2}$,
\newauthor
F. Vachier$^{2}$,
A. Marciniak$^{1}$,
J. Pollock$^{3}$,
G. Apostolovska$^{4}$,
T. Santana-Ros$^{1}$,
\newauthor
R. Hirsch$^{1}$,
W. Dimitrow$^{1}$,
M. Murawiecka$^{7}$,
P. Wietrzycka$^{1}$,
and J. Nadolny$^{5,6}$
\\
$^{1}$Astronomical Observatory Institute, Faculty of Physics, Adam Mickiewicz University, S{\l}oneczna 36, 60-286 Pozna{\'n}, Poland\\
$^{2}$IMCCE, Observatoire de Paris, Av. Denfert-Rochereau 77, 75014 Paris, France\\
$^{3}$Physics and Astronomy Department, Appalachian State University, Boone, NC 28608, USA\\
$^{4}$Institute of Physics, Faculty of Natural Sciences and Mathematics, Ss. Cyril and Methodius University, Arhimedova 3, 1000 Skopje, Macedonia\\ $^{5}$Instituto de Astrof\'isica de Canarias (IAC), E-38205 La Laguna, Tenerife, Spain\\
$^{6}$Departamento de Astrof\'isica, Universidad de La Laguna (ULL), E-38206 La Laguna, Tenerife, Spain\\
$^{7}$NaXys, Department of Mathematics, University of Namur, 8 Rempart de la Vierge, 5000 Namur, Belgium.
}

\pubyear{2016}

\begin{document}
\label{firstpage}

\date{Accepted XXX. Received YYY; in original from 2016 Dec 8}

\pagerange{\pageref{firstpage}--\pageref{lastpage}}
\maketitle

\begin{abstract}

We present a new non-convex model of the binary asteroid (809) Lundia. A SAGE
(Shaping Asteroids with Genetic Evolution) method using disc-integrated
photometry only was used for deriving physical parameters of this binary system.
The model of (809) Lundia improves former system's pole solution and gives the
ecliptic coordinates of the orbit pole -- $\lambda=122^{\circ}$,
$\beta=22^{\circ}$, $\sigma=\pm5^{\circ}$ -- and the orbital period of $15.41574
\pm 0.00001$~h.  For scaling our results we used effective diameter  of $D_{eff}
= 9.6 \pm 1.1$ km obtained from Spitzer observations. The non-convex shape
description of the components permitted a refined calculation of the components'
volumes, leading to a density estimation of $2.5\pm0.2$ g/cm$^3$ and
macroporosity of 13-23\%.  The intermediate-scale features of the model may also
offer new clues on the components' origin and evolution.

\end{abstract}


\section{Introduction}

(809) Lundia was classified as a V-type asteroid in the Flora dynamical family
\citep{flo}. The discovery of its binary nature in September 2005 was based on
photometric observations carried out at Borowiec observatory \citep{kry1}. The
first modelling of the Lundia synchronous binary system was based on 23
lightcurves obtained at Borowiec and Pic du Midi Observatories during two
oppositions in 2005/2006 and 2006/2007. The two methods of modelling - modified
Roche ellipsoids and kinematic - gave similar parameters of the system
\citep{kry2}.  The poles of the orbit in the ecliptic coordinates found  were:
longitude $118 \pm 2^\circ$, and latitude $28 \pm 2^\circ$ in the modified Roche
model and $120 \pm 2^\circ$, $18 \pm 2^\circ$ respectively in the kinematic
model \citep{kry2}.  The orbital period obtained from the lightcurve analysis as
well as from modelling was $15.418 \pm 0.001~h$. The obtained bulk density of
both components was $1.64$ or $1.71~g/cm^3$.  The comparison with HED meteorites
gave very high macroporosity of $42-49$\%.  Spectroscopic observations of the
(809) Lundia system were performed using NASA Infrared Telescope Facility and
SpeX spectrograph in 2005, 2007 and 2010. One of these spectra, observed during
total eclipse of the components allowed to investigate homogeneity/heterogeneity
of both bodies. Detailed analysis of all spectra confirmed similar mineralogy of
both components. By applying different mineralogical models a composition
similar to the one of howardite-diogenite meteorites was found \citep{birlan}.


Finding bulk density is the main goal of modelling asteroids' physical
properties, as it gives insights into their internal structure and composition.
Determining the latter based on spectroscopy or albedo probing only the surface
of the body leads to strongly biased bulk densities, but one can use that data
to find meteorite analogs for further comparison. Besides density, macroporosity
is another valuable information putting constrains on e.g. evolution or
collisional lifetime \citep{Britt2002}. Macroporosity can be found by comparing
meteorite analogs' density and porosity with the density derived from
independent method and data like photometry based shape modelling.

Density estimation relies on the estimations of the mass and volume. As
summarized by \cite{Carry2012b}, mass can be estimated using several methods:
orbit deflection during close encounters, planetary ephemeris, spacecraft
tracking or studying orbit of a satellite. Although spacecraft tracing technique
is the most precise, it is limited to small number of space missions' targets.
Very good results (of 10-15\% accuracy) can be achieved with binary
asteroids, where one can derive total mass of the system from Kepler's 3$^{rd}$
law once the satellite's orbit is known.


To know the asteroid's volume, detailed shape model and its size are needed.
Among the methods of size estimation, about 85\% of asteroids' sizes were
obtained with thermal modeling. Majority of those estimates have relative
uncertainty of about 5\%, however there are some indications of these values
being underestimated \citep{Carry2012b}. Methods capable of deriving convex
shapes only (e.g. spheres, tri-axial ellipsoids or Roche ellipsoids) put lower
constraints on the density, as introducing concavities will increase density of
the body with the same equivalent sphere diameter and period. \textit{In-situ}
observations revealed concave nature of asteroids in general, which makes
non-convex methods more adequate and accurate.

According to a list of binaries' parameters
\footnote{http://www.asu.cas.cz/~asteroid/binastdata.htm}
described first in \citep{Pravec2007}, there are only 12 double synchronous
asteroids discovered so far.
Synchronous binary asteroids with circular orbits are special cases of binary
systems where rotational periods of both components are equal to orbital one.
Additionally, angular momenta of components, orbit and resulting system's one
are parallel. This reduces the amount of free parameters of the model and allow
for detailed shape modelling leading to more accurate volume estimates.

In this paper we present a new non-convex model of the (809) Lundia system using
SAGE  (Shaping Asteroids with Genetic Evolution) method using disc-integrated
photometry only, described by \cite{bar}. This method was successfully applied
to model the binary asteroid (90) Antiope.

\section{Observations}

We continued observations of (809) Lundia system in 2007, 2008, 2009/2010, 2011,
and 2012 oppositions at Borowiec, Pic du Midi, PROMPT, South African
Astronomical Observatory, and Bulgarian National Observatory Rozhen. As
predicted the well visible eclipses/occultation events were observed only in
2011. Signs of partial eclipses/occultation are visible in the lightcurves from
2012 opposition. In Fig. 1 we show positions of the Earth in the reference frame
of the asteroid. Blue dots represent observations with eclipse/occultation
events and green without events. Open circles represent future observing
geometries and show that in 2018 only there will be a chance to observe
eclipses/occultation events.

Observations at Borowiec observatory were carried
out with $0.4m$ Newton telescope equipped with the KAF402ME CCD camera and
clear filter. The details of the Borowiec system were decsribed by
\cite{mich04}.  Observations from SAAO were carried out at $0.76m$ reflector
equipped with the University of Cape Town (UCT) CCD camera and R filter.

All CCD frames from Borowiec and SAAO were reduced for bias, dark current, and
flatfield using CCLR STARLINK package. The aperture photometry was performed to
measure the instrumental brightness of the asteroid and the comparison and check
stars.  Lightcurves observed at Pic du Midi in 2009 were obtained using $1.05m$
Cassegrain telescope equipped with THX 7863 CCD camera and L filter. Lightcurves
from 2011 were taken using Andor iKon-L CCD camera and L filter. A standard
reduction was performed using Audela software (http://www.audela.org) whose
photometry analysis was developed at IMCCE in Paris.  Observations at the
Bulgarian National Observatory at Rozhen were carried out with the $0.5/0.7m$
Schmidt telescope and KAF1602E CCD camera and R filter. For the data reduction
and aperture photometry the IDL software was used.  Three lightcurves were
obtained by PROMPT 4, a $0.41m$ Ritchey-Chr{\'e}tien telescope located in Cerro
Tololo Inter-American Observatory in Chile, equipped with Alta U47+ camera.
The aspect data of Lundia are listed in Table 1.


\begin{table*}
\caption[]{Aspect data. Columns give dates
of observations with respect to the middle of the lightcurve, asteroid's
distances to the Sun ($r$) and Earth ($\Delta$) in AU, phase angle ($\alpha$),
ecliptic longitude ($\lambda$) and latitude ($\beta$) for J2000.0 and the
observatory.}
\label{AspDat}
\begin{small}
\begin{center}
\begin{tabular}{l c c c c c c}
\hline
              &   $r$  & $\Delta$ & Phase   & $\lambda$  & $\beta$  & Observatory \\
Date (UT)     &        &          & angle   & \multicolumn{2}{c}{(J2000)}& \\
              &  (AU)  &   (AU)   &($\degr$)& ($\degr$)  &($\degr$) &  \\
\hline
 2007 Apr 04.91  &  2.7186 &  1.9977 &  17.29 & 159.95 &  3.73  & NAO Rozhen \\
 2008 May 07.32  &  2.1530 &  1.5361 &  25.38 & 292.88 &  9.27  & PROMPT \\
 2008 May 08.29  &  2.1509 &  1.5242 &  25.25 & 293.05 &  9.31  & PROMPT \\
 2008 May 09.30  &  2.1487 &  1.5118 &  25.11 & 293.22 &  9.36  & PROMPT \\
 2008 Jun 10.92  &  2.0776 &  1.1682 &  16.68 & 294.94 & 10.59  & SAAO \\
 2009 Nov 20.10  &  2.4282 &  1.9806 &  23.26 & 133.39 & -6.17  & Pic du Midi \\
 2009 Nov 25.14  &  2.4336 &  1.9276 &  22.57 & 133.93 & -6.23  & Pic du Midi \\
 2011 Apr 07.07  &  2.5479 &  1.6394 &  11.81 & 226.89 &  9.72  & Pic du Midi \\
 2011 Apr 08.10  &  2.5462 &  1.6310 &  11.86 & 226.71 &  9.79  & Pic du Midi \\
 2011 Apr 09.06  &  2.5446 &  1.6334 &  11.09 & 226.55 &  9.85  & Pic du Midi \\
 2011 Apr 11.10  &  2.5414 &  1.6080 &  10.34 & 226.17 &  9.97  & Pic du Midi \\
 2011 Apr 12.08  &  2.5399 &  1.6009 &   9.98 & 225.98 &  10.03 & Pic du Midi \\
 2012 Oct 09.04  &  1.9847 &  1.2466 &  24.71 &  71.60 & -10.87 & Borowiec  \\
 2012 Oct 18.05  &  2.0017 &  1.1901 &  21.59 &  71.60 & -11.69 & Borowiec  \\
 2012 Oct 18.02  &  2.0035 &  1.1846 &  21.21 &  71.55 & -11.77 & Borowiec  \\
 2012 Oct 20.07  &  2.0056 &  1.1787 &  20.80 &  71.49 & -11.86 & Borowiec  \\
 2012 Nov 11.05  &  2.0495 &  1.1012 &  10.87 &  67.95 & -13.31 & Borowiec  \\
 2012 Nov 26.03  &  2.0810 &  1.1086 &   6.36 &  63.95 & -13.50 & Borowiec  \\
\hline
\end{tabular}
\end{center}
\end{small}
\end{table*}

\begin{figure}

\centering

\includegraphics[width=80mm]{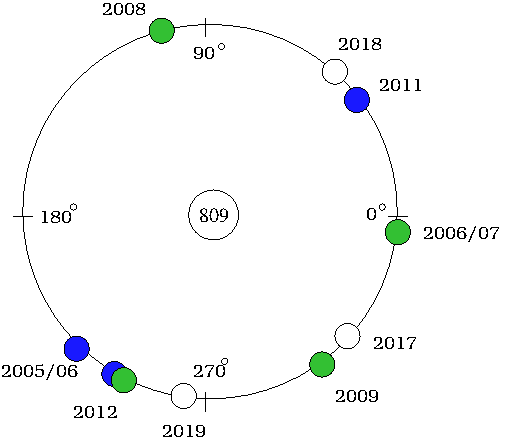}

\caption{Positions of the Earth in the reference frame of the asteroid. Blue dots
represent positions with observed eclipse/occultation events and green without events.
Open circles represent future observing geometries.}


\end{figure}

The obtained composite lightcurves are presented in Figs. 2-6. Lighcurves are
composed with the synodic period of $15.418 \pm 0.001~h$. For good comparison
between lightcurves the scale on each graph is the same.  In Fig 2. we present a
lightcurve from NAO Rozhen in comparison with already published lightcurves from
Pic du Midi. Despite the time span between the lightcurves is as much as four
months the internal fit is very good and it confirms the derived synodic period.
Currently, our dataset consists of 41 individual lightcurves obtained during 6
oppositions and covering different observing geometries. Details of the
observing geometries of the (809) Lundia system for each opposition are given in
Table 2.


\begin{figure}
\label{fig:2007}
\centering
\includegraphics[width=8.5cm]{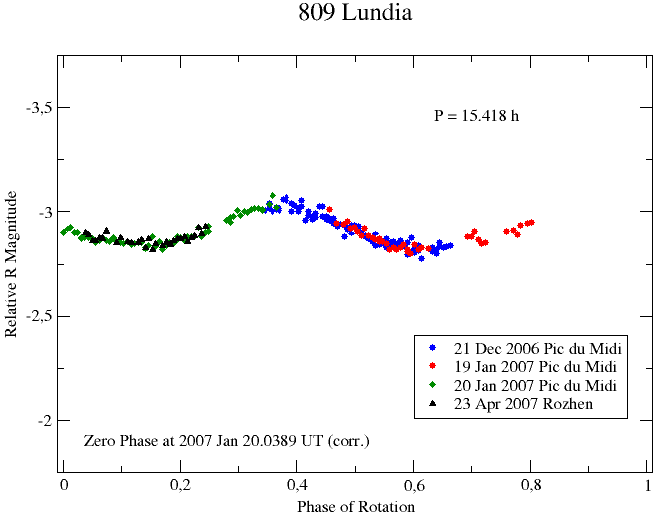}
\includegraphics[width=8.5cm]{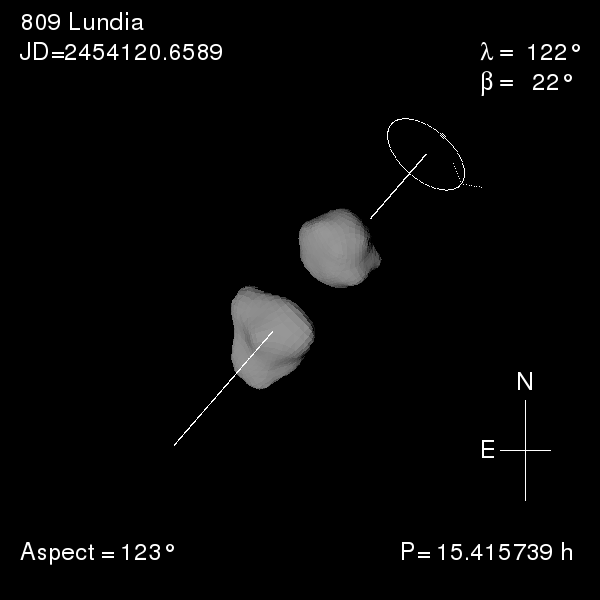}
\caption{Composite lightcurves of 809 Lundia from 2006/2007 opposition (top)
with corresponding view of the system from Earth (bottom).}
\end{figure}

\begin{figure}
\label{fig:2008}
\centering
\includegraphics[width=8.5cm]{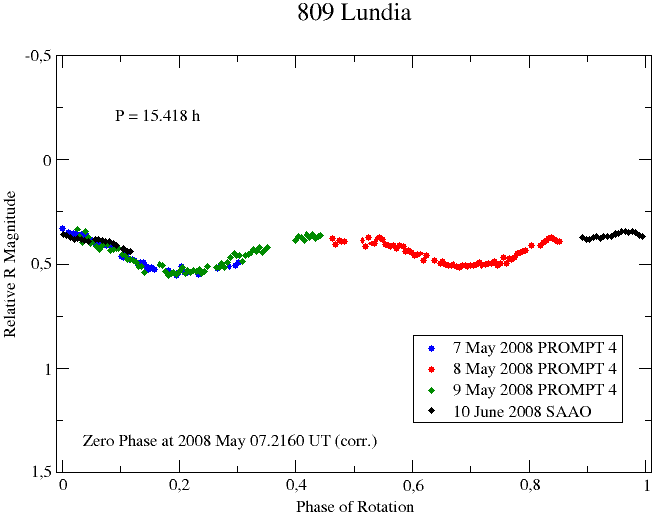}
\includegraphics[width=8.5cm]{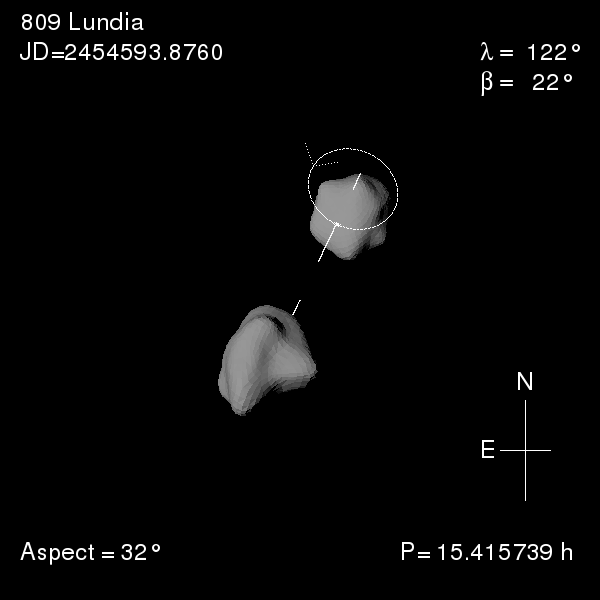}
\caption{Composite lightcurves of 809 Lundia from 2008 opposition (top)
with corresponding view of the system from Earth (bottom).}
\end{figure}

\begin{figure}
\label{fig:2009}
\centering
\includegraphics[width=8.5cm]{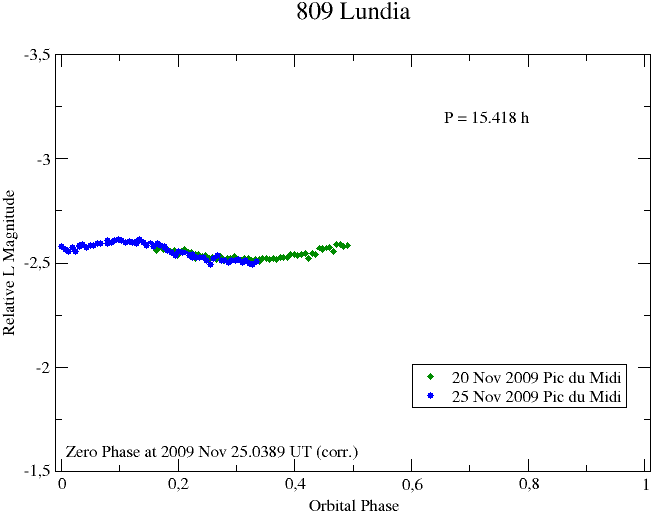}
\includegraphics[width=8.5cm]{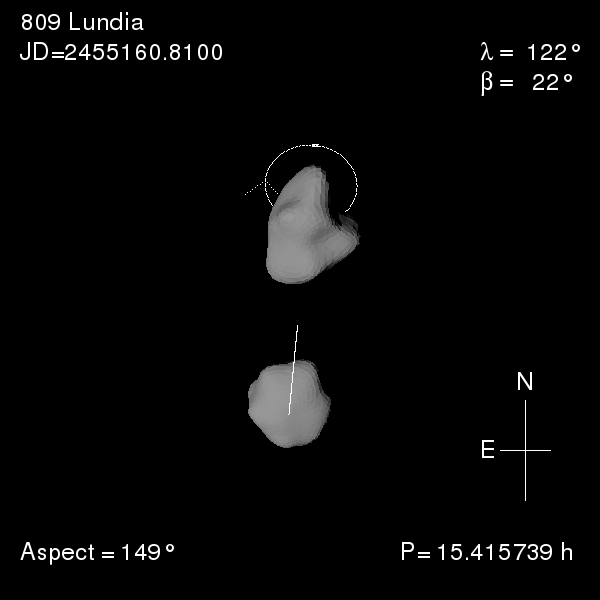}
\caption{Composite lightcurves of 809 Lundia from 2009 opposition (top)
with corresponding view of the system from Earth (bottom). }
\end{figure}

\begin{figure}
\label{fig:2011}
\centering
\includegraphics[width=8.5cm]{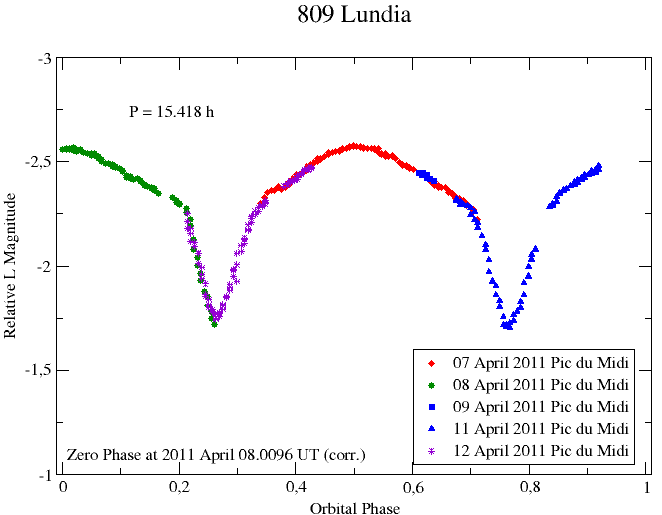}
\includegraphics[width=8.5cm]{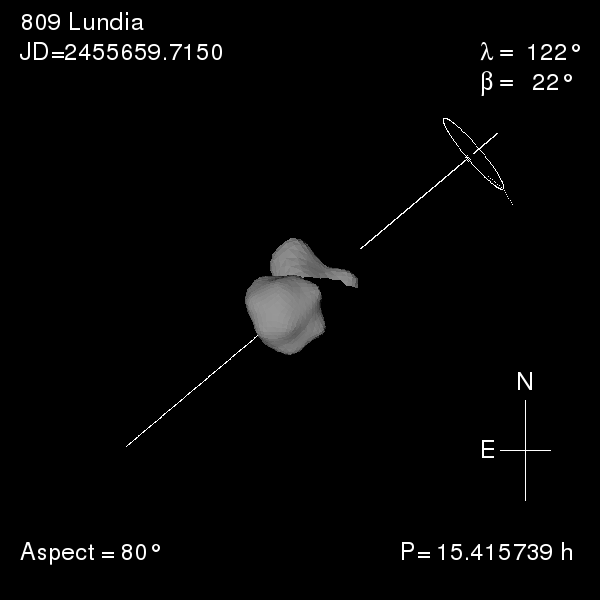}
\caption{Composite lightcurves of 809 Lundia from 2011 opposition (top)
with corresponding view of the system from Earth (bottom). }
\end{figure}

\begin{figure}
\label{fig:2012}
\centering
\includegraphics[width=8.5cm]{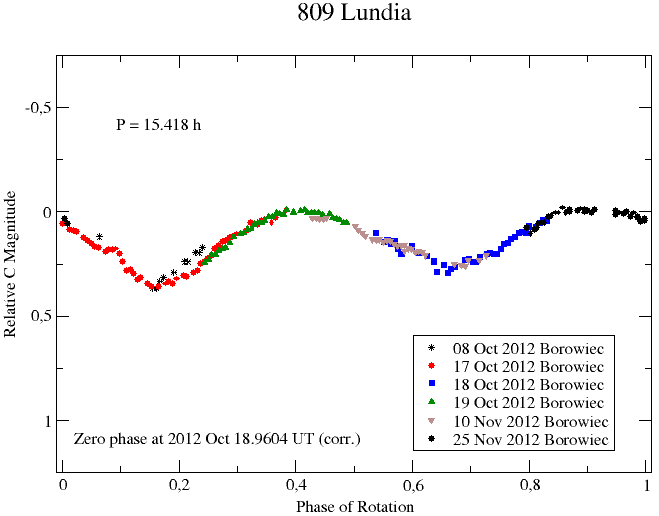}
\includegraphics[width=8.5cm]{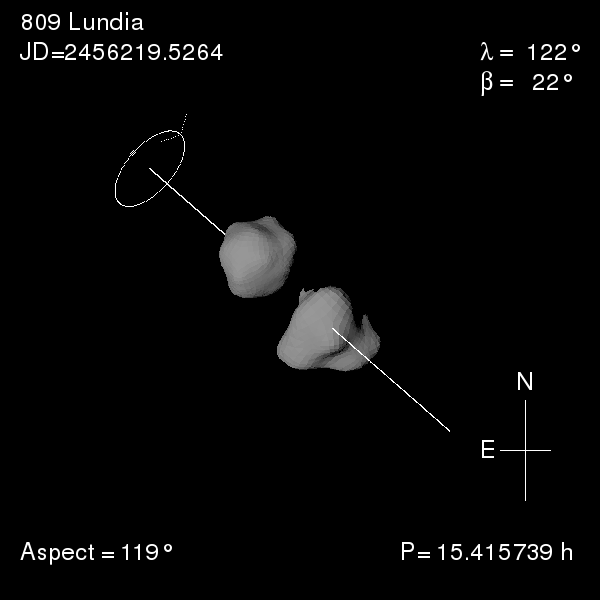}
\caption{Composite lightcurve of 809 Lundia from 2012 opposition (top)
with corresponding view of the system from Earth (bottom). }
\end{figure}

\begin{table*}

\begin{minipage}{170mm}

\caption{
	Details of the lightcurves used for our modelling of (809) Lundia The table
columns describe the observations time range, the number of observing nights,
the phase angle range, the ecliptic longitudes and latitudes of the asteroid
around the opposition dates, the observed eclipsing amplitudes and the
references. The value of '0' for the eclipsing amplitudes means that no eclipses
were observed for these apparitions (i.e. 2006/2007, 2008 and 2009).}

  \label{observations}

\begin{tabular}{cccccccc}

\hline

Time range & $N_\mathrm{lc}$ & $\alpha$ $(^{\circ})$ & $\lambda$ $(^{\circ})$ & $\beta$ $(^{\circ})$ & Eclipsing  & Reference\\

  &  &  &  &  & amplitude (mag) &\\

\hline

 Sep 2005 -- Jan 2006 & 19  &  7.3 -- 25.9 & 45  & -11 & 0.75 -- 0.30 & \citet{kry1}\\

 Dec 2006 -- Apr 2007 & 4+1 & 17.3 -- 21.4 & 168 &   2 & 0 & \citet{kry1} and this paper \\

 May 2008 -- Jun 2008 & 4   & 16.7 -- 25.3 & 294 &  10 & 0 & this paper \\

 Nov 2009             & 2   & 22.6 -- 23.3 & 133 &  -6 & 0 & this paper \\

 Apr 2011             & 5   & 10.0 -- 11.8 & 226 &  10 & 0.58 & this paper \\

 Oct 2012 -- Nov 2012 & 6   &  6.4 -- 24.7 & 68  & -12 & 0 -- 0.15 & this paper \\

\hline

\end{tabular}

\end{minipage}

\end{table*}

\section{Method}

We used genetic-algorithm-based modelling method SAGE that given solely
photometric observations recreates non-convex shape, spin axis orientation and
rotational period of synchronous binary asteroid \citep{bar}.

The asteroid system is assumed to be synchronous with circular orbit about
center of mass. Each of the bodies is described by 62 vectors with fixed
directions; the length of each vector is a free parameter during modelling
process. To create lightcurves of the system a refined and more detailed
shapes are used, created using surface smoothening algorithm
\citep{Catmull-Clark}. To fully describe the system there are two additional
free parameters: bodies size ratio and separation. During modelling process the
best fit for orbital period is searched and than used to establish mass ratio
using the separation, shapes and size ratio as well.

In every step of modelling process synthetic lightcurves are compared with
observed ones. To calculate a lightcurve for specific moment of time, a geometry
of the observations (i.e. the positions of the Sun, asteroid and Earth) is
reconstructed using position vectors in heliocentric, ecliptic reference frame
based on an ephemeris of the asteroid. When a 3D scene is constructed a system
is rotated $360^{\circ}$ to create a full rotation lightcurve that is later used
for comparison. The flux of an asteroid is obtained in rasterisation process:
the image is  created as if a telescope with an infinite resolution observed an
asteroid creating a CCD image. The sum of the pixels gives a synthetic relative
photometry measurement. A Lommel-Seeliger scattering law is used and no albedo
variations are assumed.

The modelling process uses genetic algorithm to arrive at global minimum. The
modelling starts with the two spheres, random spin axis orientation and
approximate rotational period. In every step a random changes to the shapes,
period, spin axis orientation, separation and size ratio are applied creating a
random population (generation) of systems. Each body of the system is a physical
model assuming homogeneous distribution of mass; the axes of largest inertia of
the bodies are aligned with spin axis of the whole system. Than, for every model
of the system in the population a synthetic lightcurves are calculated and
compared with observations using $\chi^2$ test. The model with smallest $\chi^2$
is chosen as the seed for the next population and the whole process repeats
until $\chi^2$ value no longer changes from population to population. In order
to assure the modelling process not to fall into local minimum a weighting
process is applied in every step. The lightcurve with largest $\chi^2$ is given
the largest weight to steer the process towards the global minimum; the weights
change in every step.

Additionally, the whole modelling process is run multiple times creating a
family of solutions. This is a standard procedure in genetic evolution
algorithms to ensure the result being in global, rather than local, minimum. The
path leading to a model in each run is different, but the models should be alike
in the end. If the models differ significantly, it indicates that not enough
observational data has been supplied for the modelling.

\section{Model of (809) Lundia}

\begin{figure*}
	\label{model1}
	\centering
	\includegraphics[width=16cm]{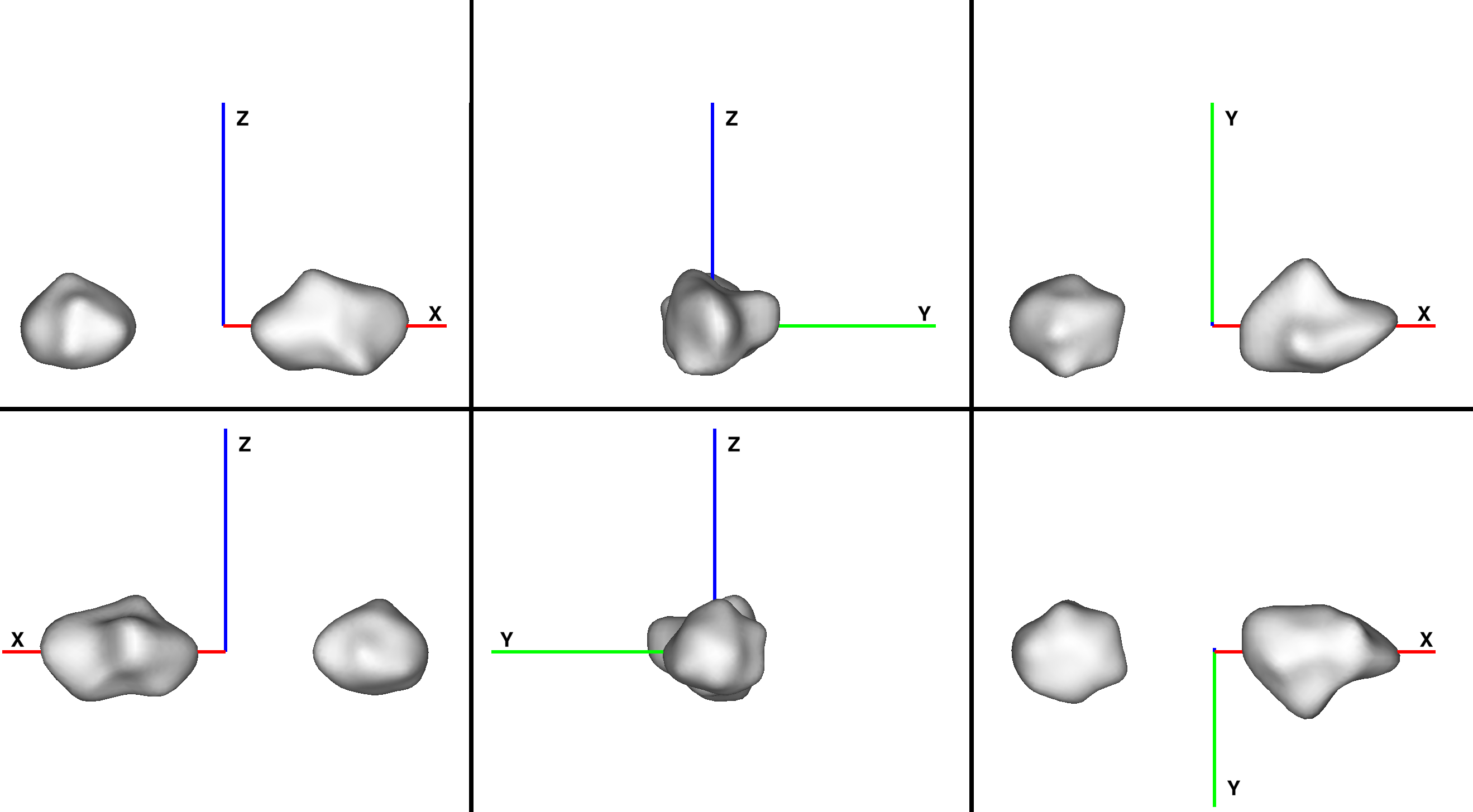}
	\caption{$xz$, $yz$, $xy$, $-xz$, $-yz$, $-xy$ projections of the Lundia
		model.}
\end{figure*}

\begin{figure*}
	\label{lc_comparison}
	\centering
	\includegraphics[width=\textwidth]{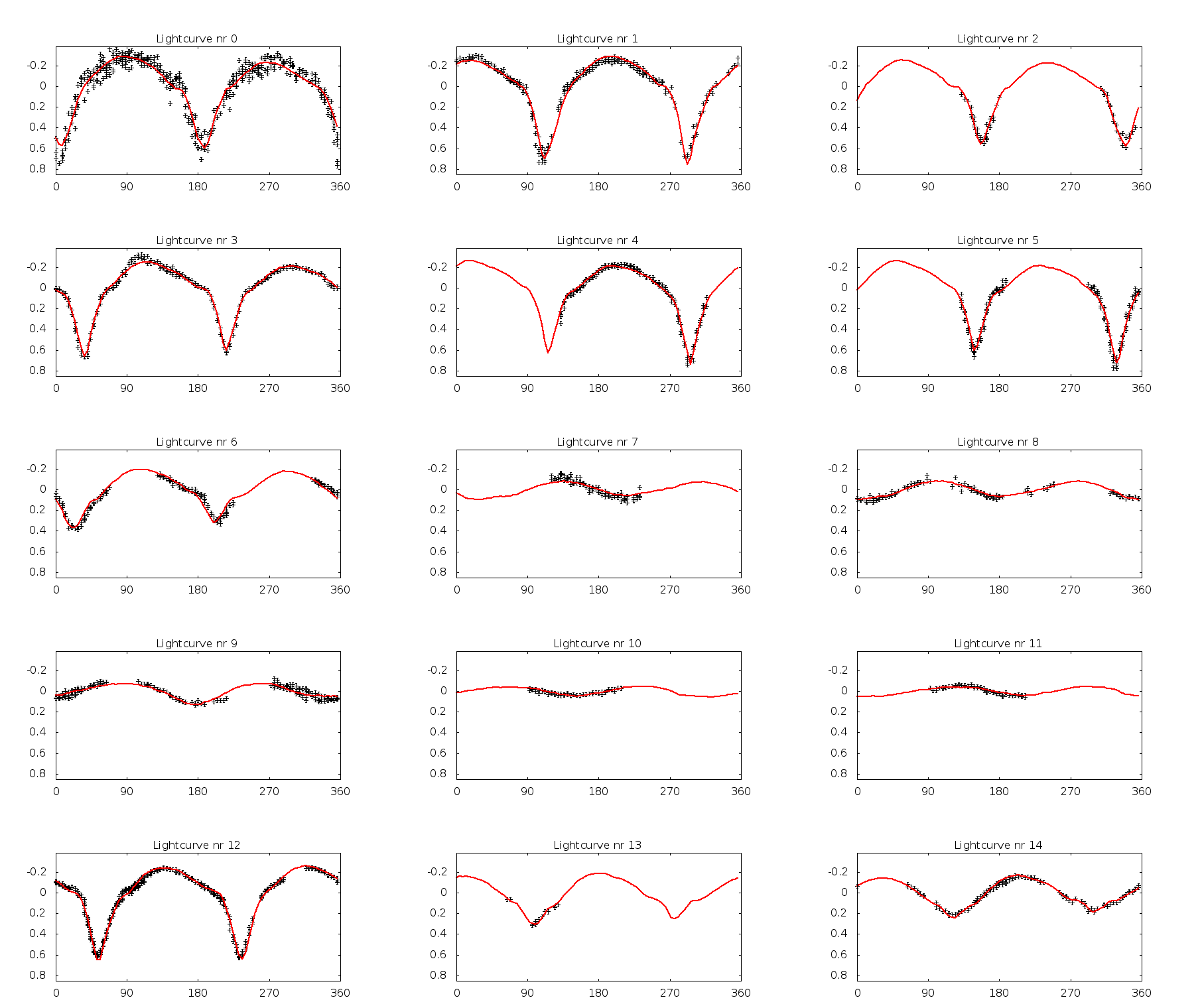}
	\caption{Observations (black crosses) versus synthetic lightcurves of the
		model (solid red line).}
\end{figure*}

The Lundia shape model projections can be seen in Fig.~\ref{model1}. Synthetic
lightcurves generated by the model fit the observed ones very well
(Fig.~\ref{lc_comparison}). Mutual eclipse events are perfectly timed and the
lightcurves' shape is reproduced on general and detail levels. The uncertainty
of photometry was not reported by the observers and therefore it is assumed to
be $0.02~mag$.

Using Spitzer Space telescope \cite{marchis} calculated effective diameter of
Lundia $D_{eff}=9.6~km$ with $3\sigma$ uncertainty of $1.1~km$. Combining this
result with inhomogeneous Roche ellipsoids model by \cite{descamps2010} they
obtained equivalent sphere diameters for the primary and secondary components
$D_p=7.2\pm1.4~km$ and $D_s=6.4\pm1.3~km$, with system separation $d=14.6~km$ and
average bulk density of the system $\rho=1.77~g/cm^3$.

Applying said effective diameter we scaled the new non-convex Lundia model by
assuming the same volume of the system. Mass was determined from the orbital
period $P$ and system separation $d$ using Kepler's third law.

The parameters of non-convex model of Lundia system are:
\begin{itemize}
	\item system's spin axis ecliptic coordinates:
		\begin{itemize}
			\item[] $\lambda=122.5 \pm 5\st$,
			\item[] $\beta=22 \pm 5\st$
		\end{itemize}
	\item sidereal period: $P=15.41574 \pm 10^{-5}~h$
	\item primary equivalent sphere diameter: $D_p=8.1 \pm 0.9~km$
	\item secondary equivalent sphere diameter: $D_s=7.1 \pm 0.8~km$
	\item $D_s/D_p = 0.87$
	\item system separation: $d=18.2 \pm 0.8~km$
	\item total mass: $(1.157 \pm 0.4) 10^{15}~kg$
	\item bulk density: $\rho=2.5 \pm 0.2~g/cm^3$
	\item macroporosity: 13-23\%
\end{itemize}

Spectroscopic observations' analysis \citep{birlan} indicates similar
mineralogical composition as howardite-diogenite meteorites.  The obtained
density of $2.5~g/cm^3$ is much higher than determined before, and this value in
comparison with the density of HED meteorites of $2.86$ to $3.26~g/cm^3$
reported by \cite{britt} and \cite{McCau} infers the macroporosity of (809)
Lundia of only 13-23\%, rather than 40-50\% as reported by \cite{marchis}.

%
%
%

\section{Conclusions}

The obtained non-convex model of (809) Lundia perfectly reproduces the obtained
set of photometric lightcurves. By scaling volume of the system components we
found density 50\% higher than the one calculated in the previous studies
and macroporosity of about 60\% smaller than before \cite{kry2}. However the
internal structure of asteroids is presently unclear and
newly obtained values may serve as an input to the theories of the Solar
System evolution.
Higher values of the bulk density are due to non-convex shape of the system. The
more shape deviates from a sphere the larger density value we get.

The shape and size of Lundia could be refined given direct measurements, like
stellar occultation events. Predictions for 2017 based on new GAIA DR1 catalog
yield two events on 27 April and 28 May 2017. Unfortunately Lundia's small size
makes accurate prediction difficult, as uncertainty of star position on the
level of $7$ $mas$ translates into $15$ km uncertainty in the position on Earth
which is about the size of the Lundia system. Nevertheless, such observations
could put better constrains on asteroid's size and density.

\section*{Acknowledgements}

The research leading to these results has received funding from the European
Union's Horizon 2020 Research and Innovation Programme, under Grant Agreement no
687378.

This work was partialy supported by grant no. 2014/13/D/ST9/01818 from the
National Science Centre, Poland

This paper uses observations made at the South African Astronomical Observatory
(SAAO).  The reduction of CCD frames from Borowiec and SAAO were performed with
the CCLR STARLINK package.

G. Apostolovska gratefully acknowledge observing grant support from the
Institute of Astronomy and Rozhen National Astronomical Observatory, Bulgarian
Academy of Sciences



\bibliographystyle{mnras}
\bibliography{Bartczak_Lundia} 

\bsp    
\label{lastpage}
\end{document}